\documentclass[fleqn,usenatbib]{mnras}

\usepackage[T1]{fontenc}

\usepackage{float}
\DeclareRobustCommand{\VAN}[3]{#2}
\let\VANthebibliography\thebibliography
\def\thebibliography{\DeclareRobustCommand{\VAN}[3]{##3}\VANthebibliography}

\usepackage{graphicx}	
\usepackage{amsmath}	
\usepackage{amssymb}	
\usepackage{subcaption}

\newcommand{\hGpc}{\,h^{-1}~{\rm Gpc}}

\newcommand{\hMpc}{{\ifmmode{\,h^{-1}{\rm Mpc}}\else{$h^{-1}$Mpc}\fi}}
\newcommand{\hkpc}{{\ifmmode{\,h^{-1}{\rm kpc}}\else{$h^{-1}$kpc}\fi}}
\newcommand{\hMsun}{{\ifmmode{\,h^{-1}{\rm {M_{\odot}}}}\else{$h^{-1}{\rm{M_{\odot}}}$}\fi}}

\newcommand{\Mstar}{{\ifmmode{\,M_{*}}\else{$M_{*}$}\fi}}
\newcommand{\Mhalo}{{\ifmmode{\,M_{\rm halo}}\else{$M_{\rm halo}$}\fi}}
\newcommand{\ltsima}{$\; \buildrel < \over \sim \;$}
\newcommand{\gtsima}{$\; \buildrel > \over \sim \;$}
\newcommand{\lsim}{\lower.5ex\hbox{\ltsima}}
\newcommand{\gsim}{\lower.5ex\hbox{\gtsima}}

\newcommand{\theth}{{\sc The Three Hundred}}
\newcommand{\ahf}{\textsc{AHF}}

\newcommand{\gadgetx}{\textsc{Gadget-X}}
\newcommand{\simba}{\textsc{GIZMO-SIMBA}}
\newcommand{\gadgetmusic}{\textsc{Gadget-MUSIC}}

\title[Cluster Dynamical States]{\theth: Cluster Dynamical States and Relaxation Time Scale}

\author[Zhang et al.]{Bowei Zhang,$^{1,2}$\thanks{E-mail: bz287@cam.ac.uk (BZ)}
Weiguang Cui,$^{1}$\thanks{E-mail: weiguang.cui@ed.ac.uk (WC)}
Romeel Dave,$^{1}$ and 
Marco De Petris$^{3}$
\\
$^{1}$Institute for Astronomy, University of Edinburgh, Royal Observatory, Edinburgh EH9 3HJ, United Kingdom\\
$^{2}$Department of Applied Mathematics and Theoretical Physics, University of Cambridge,Cambridge CB3 0WA, United Kingdom\\
$^{3}$Dipartimento di Fisica, Sapienza Universit\'a di Roma, Piazzale Aldo Moro, 5-00185 Roma, Italy
}

\date{Accepted XXX. Received YYY; in original form ZZZ}

\pubyear{2021}

\begin{document}
\label{firstpage}
\pagerange{\pageref{firstpage}--\pageref{lastpage}}
\maketitle

\begin{abstract}
Using the galaxy clusters from \theth, we define a new parameter: $\lambda_{DS}$ to describe the dynamical state of clusters, which assumes a double-Gaussian distribution in logarithm scale for our mass-complete cluster sample at $z=0$ from the dark-matter-only (DMO) run. Therefore, the threshold for distinguishing relaxed and unrelaxed clusters is naturally determined by the crossing point of the double-Gaussian fitting which has a value of $\lambda_{DS} = 3.424$. By applying $\lambda_{DS}$ with the same parameters from the DMO run to the hydro-dynamically simulated clusters (\gadgetx\ run and \simba\ run), we investigate the effect of baryons on the cluster dynamical state. We find a weak baryon-model dependence for the $\lambda_{DS}$ parameter. Finally, we study the evolution of $\lambda_{DS}$ along with clusters mass accretion history. We notice an upper limit of halo mass change $\frac{\Delta M_{200}}{M_{200}} \sim 0.12$ that don't alter the cluster dynamical state, i.e. from relaxed to unrelaxed. We define relaxation period (from the most relaxed state to disturb and relaxed again) which reflects how long the dynamical state of a cluster restores its relaxation state, and propose a correlation between this relaxation period and the strength of halo mass change $\frac{\Delta M_{200}}{M_{200}}$. With the proposed fitting to such correlation, we verify the relaxation period can be estimated from $\frac{\Delta M_{200}}{M_{200}}$ (including multi mass change peaks) with considerably small error.
\end{abstract}




\section{Introduction}
Galaxy clusters are the largest gravitationally bounded structures in the Universe. Masses of clusters, ranging from $\sim10^{14}M_{\odot}$ to over $10^{15}M_{\odot}$, are dominated by dark matter ($\sim$80$\%$ - 85\%) \citep[e.g.][]{White1993,Fukugita1998}. The gravitational processes exhibiting by the dynamical properties of dark matter halo can be viewed from the baryonic compositions in galaxy clusters, such as the galaxies and intra-cluster medium (ICM), which emit photons at different wavelengths and thus can be observed through telescopes. As the "brightest" objects in the sky, especially in X-ray band, the galaxy clusters, such as the Coma cluster, used to be the main targets of astronomical observation. Therefore, numerous studies have focus on the cluster properties. Cluster dynamical state -- one unique feature to describe the cluster virialized state, connects to many other important cluster quantities such as its mass \citep[e.g.][]{Nelson2012,Biffi2016,Gianfagna2021}, formation time \citep[e.g.][]{300Mostoghiu2019}, and concentration \citep[e.g.][]{Neto:2007vq}, etc.

Galaxy clusters mergers, the most powerful events which violate the cluster equilibrium, provide unique conditions to study a range of physics \citep[e.g.][]{poole_impact_2006,zenteno_joint_2020}. From the perspective of cosmology, disturbed clusters provide test bed for $\Lambda$CDM model \citep[e.g.][]{Thompson:2014zra,Kim:2016ujt,Sereno2018}, and their enhanced strong lensing efficiency provide powerful tools to investigate the universe at high redshift \citep[see][for example]{Baldi:2012mp,Acebron:2018mac}. 

From the perspective of galaxy evolution, cluster dynamical state is also valuable. For example, \cite{morell_classification_2020} found that galaxies evolve in the same way in Gaussian and Non-Gaussian systems, but their formation histories leads to different mixtures of galactic types and infall patterns. Furthermore, disturbed clusters can be used to examine the effect of the ram pressure from the cluster ICM thanks to these jellyfish galaxies inside \citep{McPartland2015}. Lastly, we found that the halo formation time affect the central BCG properties (Cui et al. in prep, see \citet{Cui2021} for a similar result at lower halo mass), thus this can be also linked to the cluster dynamical state. 

At the late stage of the hierarchical structure formation process, the recent formation history of massive structures such as galaxy clusters, correlates strongly with the degree of its dynamical equilibrium \citep[e.g.][]{Wong2012}. This is easy to understand as a recent major merger will result the cluster in dynamical disequilibrium for a certain time, while earlier major merger means a following decreased mass accretion. As a result of those highly complicated dynamical processes of merger events, galaxy clusters can have a wide range dynamical states, which can be reduced to two categories: relaxed(or virialized) and non-relaxed(or non-virialized). Relaxed clusters are expected to have nearly spherical shape and Gaussian velocity distribution \citep[e.g.][]{Faltenbacher:2006rb}, while non-relaxed clusters can show elongated shapes \citep{Gouin2021}, non-Gaussian velocity distribution \citep{Hou:2009fi}, the presence of massive substructures \citep{Lopes:2018ysc} and irregular morphology properties \citep[e.g.][]{300DeLuca2021}.

With the purpose of better utilization of galaxy clusters as probes of cosmology and testbeds for galaxy formation and evolution, it is of great importance to interpret the dynamical processes in galaxy clusters. Different methods for classifying cluster dynamical states and assessing the relaxation degree of clusters are found in the literature \citep[see][and references therein for theoretical approaches]{cui_dynamical_2017}, \citep[see][and references therein for observational-like approaches]{300DeLuca2021}. In observations, cluster dynamical states is classified by and related to the ICM hydrostatic equilibrium: clusters that are undergoing, or have undergone a merger process, leave the ICM in turbulent. \cite{wen_substructure_2013} developed method to diagnose the substructure and dynamical state of galaxy clusters by using photometric data of Sloan Digital Sky Survey. \cite{300Capalbo2021} and \cite{300DeLuca2021} investigated the correlation between cluster dynamical states and cluster morphology, which is directly measured through images of the surface brightness in the X-ray band and of the y parameter from the thermal Sunyaev–Zel’dovich effect in the millimetre range. In theory, there are a vague of ways with which halo dynamical state can be evaluated. Using DMO simulations, \cite{Bett:2006zy} used integrated virial ratio $2T/|W|$+1 to classify dynamical states, and suggested $2T/|W|$+1 < 1.5 to select halos in quasi-equilibrium states. \cite{Neto:2007vq} expanded the criteria by including substructure mass fraction and centre-of-mass offset, which contain the information of the constituents in cluster and the shape of cluster respectively. \cite{shaw_statistics_2006} additionally took the surface pressure energy $E_s$ into account in virial ratio calculation \citep[see also][for detailed calculation for hydrodynamic simulations]{cui_dynamical_2017}. \cite{Davis:2010ei} found the effect of the potential energy from particles outside of halos is negligible. Nevertheless, all of these methods require manually set thresholds to split the cluster dynamical state into relaxed and un-relaxed (disturbed) in both theory and observation, which doesn't generally showing up in their distributions \citep[in either single or combined parameters][]{300DeLuca2021,300Haggar2020}. Therefore, our first task in this paper is to remove these thresholds and present a new parameter-free cluster dynamical state classification. 

One key question which is not well thoroughly studied in the literature is the relaxation time scale of the cluster dynamical state, i.e. how long does it need from getting disturbed to relaxed in hydrostatic equilibrium. This will help us to understand the cluster thermalisation \citep{300Sereno2021}. Furthermore, it is also interesting to see the effect of baryons on the cluster dynamical state and their evolution. 

The layout of this paper is as following: we introduce the \theth\ project in section \S\ref{s:300}. The new parameter-free cluster dynamical classification method and separation of relaxed and un-relaxed clusters are presented in \S\ref{s:dsm}. Our main results on the cluster dynamical state is shown in \S\ref{s:results}. We finally conclude and discuss our study on cluster dynamical state in \S\ref{s:conc}.


\section{The Three Hundred project}\label{s:300}

The \theth\footnote{\url{https://the300-project.org}} consists of 324 re-simulated clusters and 4 field regions extracted from the MultiDark Planck simulation, MDPL2 \citep{Klypin2016}. The MDPL2 simulation has cosmological parameters of $\Omega_M=0.307, \Omega_B=0.048, \Omega_{\Lambda}=0.693, h=0.678, \sigma_8=0.823$.
All the clusters and fields have been simulated using the full-physics hydrodynamic codes \gadgetx\ \citep[GX in short,][]{Rasia2015,Steinborn2015,Beck2016}, \gadgetmusic\ \citep{MUSICI},
which are updated versions of {\sc Gadget2} \citep{Springel2005} and \simba\ \citep[GIZMO in short,][and see Cui et al. 2021 in prep. for the details of the \theth cluster run]{Dave2019} which is developed from {\sc MUFASA} \citep{Dave2016} using the {\sc GIZMO} code \citep{Hopkins2015}.
In the re-simulation region, the mass of a dark matter particle is $12.7\times10^8 \hMsun$ and the mass of a gas particle is $2.36\times10^8 \hMsun$.
Each cluster re-simulation consists of a spherical region of radius $15 \hMpc$ at $z=0$ centred on one of the 324 largest objects within the host MDPL2 simulation box, which is $1 \hGpc$ on a side. The halo masses of central galaxy clusters range from
$\sim 6.4\times10^{14} \hMsun$ to $2.63\times{10}^{15} \hMsun$.

A more detailed introduction of \theth\ can be found in \cite{300Cui2018}. Besides these studies on the cluster dynamical state which has been mentioned in the introduction, these simulated galaxy clusters have been used for different proposes: the filaments around the clusters \citep{300Rost2021,300Kuchner2020,300Kuchner2021,300Kotecha2021}; the backsplash galaxies \citep{300Haggar2020,300Knebe2020} and shock radius \citep{300Baxter2021,Anbajagane2021}. The advanced baryon models in hydrodynamic simulations allow us to perform a detailed investigation on the cluster properties, such as profiles \citep{300Mostoghiu2019,300Li2020}, substructure and its baryonic content \citep{300Arthur2019,300Haggar2021,300Mostoghiu2021,300Mostoghiu2021b}, the cluster (non-)thermalization \citep{300Sayers2021,300Sereno2021}, the fundamental plane \citep{300Diaz-Garcia2021}, and the cluster mass bias \citep{300Ansarifard2020, 300Li2021, 300Anbajagane2021}. Additional runs allow us to investigate more things: such as the effect of environment by comparing to void/field regions \citep{300Wang2018}; constraining the dark matter cross-section with the self-interacting dark matter run \citep{300Vega-Ferrero2021}; examining the chameleon gravity \citep{300Tamosiunas2021}.

In this paper, we only use the halos identified by the Amiga's Halo Finder \citep[\ahf][]{Knollmann2009} with a spherical overdensity of 200 $\times \rho_{crit}$. The progenitors of these halos are tracked and identified using the {\sc Mergertree} that is part of the \ahf\ package. We only focus on the main progenitors of the cluster, which is defined as the highest matched halo in previous snapshot, for tracking their mass accretion history.

\section{The parameter-free classification of cluster dynamical state} \label{s:dsm}

\subsection{Dynamical parameters and previous work on classifying cluster dynamical states}

In the literature, for example, \cite{cui_dynamical_2017}, different parameters are used to describe the dynamical states of clusters. The commonly used parameters are:
\begin{itemize}
\item The Virial Ratio, $\eta$.

The exact expression for the virial theorem is 
\begin{equation}
    \frac{1}{2} \frac{d^{2}I}{dt^{2}} = 2T+W-E_{s}
    \label{E1}
\end{equation}

where $I$ is the moment of inertia, $T$ and $W$ are kinetic energy and potential energy respectively, and $E_{s}$ is the energy from surface pressure $P$.

If the cluster system is in dynamical equilibrium, the \autoref{E1} will reduce to  
\begin{equation}
    2T+W-E_{s} = 0
    \label{E2}
\end{equation}

which can be rewritten as 
\begin{equation}
    \frac{2T-E_{s}}{|W|} = 1
    \label{E3}
\end{equation}

Therefore, the virial ratio is defined as
\begin{equation}
    \eta = \frac{2T-E_{s}}{|W|}
    \label{E4}
\end{equation}

and a relaxed cluster is expected to have $\eta\approx1$.
\item Subhalo Mass Fraction, $f_{s}$.
 
$f_{s}$ represents the fraction of total mass of cluster contained in subhalos, which are identified by \ahf. However, this fraction doesn't include the most massive substructure since it only includes the bounded components in main halos.
For the most relaxed clusters, the subhalo mass fraction should be very small.

\item Center of Mass Offset, $\Delta_{r}$.

The offset of the center of mass of cluster is defined as
\begin{equation}
    \Delta_{r} = \frac{\emph{\textbf{R}}_{cm}-\emph{\textbf{R}}_{c}}{R_{vir}}
    \label{E5}
\end{equation}

where $R_{vir}$ is virial radius, within which virial theorem applies for a bounded system. $\emph{\textbf{R}}_{c}$ is the position of the peak of the density field of cluster, and $\emph{\textbf{R}}_{cm}$ is the position of the center of mass, which can be calculated by

\begin{equation}
    \emph{\textbf{R}}_{cm}= \frac{1}{M}\sum\limits_{i=1}^{n}m_{i}r_{i}
    \label{E6}
\end{equation}
 where $m_{i}$ and $r_{i}$ are the mass and position of the $i^{th}$ particle, $M$ is the virial mass of the halo and $n$ is the total number of particles within $R_{vir}$.
 
 Empirically, a gravitationally bounded system in equilibrium has symmetric mass distribution, which requires small distance between the center of mass and the peak of density. Therefore, small $\Delta_{r}$ is expected for a relaxed cluster.
\end{itemize}

We emphasise here that all the three parameters are only phenomenal descriptions\footnote{$\eta$ is the closest one to the physical definition of dynamical equilibrium with $\eta = 1$. However, the cluster can not be treated as an isolated object. Even with the surface pressure correction term, we can not fully correct $\eta$, for example, the effect of potential coming from a nearby object.} of the cluster dynamical state due to the lack of a physically defined quantity for it. Furthermore, it is not clear which parameter contribute more to or describe better the cluster dynamical state. Therefore, different criteria are applied to classify a cluster as relaxed. For example,
\cite{cui_dynamical_2017} concluded that a relaxed cluster should satisfy three criteria: $\Delta_{r} < 0.04$, $f_{s} < 0.1$ and $0.85<\eta<1.15$ . With these criteria, \cite{300Haggar2020} combined these three parameters which are normalised to their thresholds but with equal weight, to a continuous, non-binary measure of cluster dynamical states, which is defined as the "relaxation" parameter of cluster, $\chi_{DS}$:

\begin{equation}
    \chi_{DS}= \sqrt{\frac{3}{(\frac{f_{s}}{0.1})^{2}+(\frac{|1-\eta|}{0.15})^{2}+(\frac{\Delta_{r}}{0.04})^{2}}}
    \label{E7}
\end{equation}

For a cluster to be dynamically relaxed, it requires $\Delta_{r}$ and $f_{s}$ to be minimised, and $\eta \approx 1$. Therefore, the most relaxed clusters are expected to have large $\chi_{DS}$ ($\chi_{DS}$ > 1). However, there is no clear separation between the relaxed and unrelaxed clusters. Actually, the distribution of the $\chi_{DS}$ exhibits a single peak Gaussian curve. Therefore, we need to manually set a threshold to break the cluster dynamical state into relaxed and unrelaxed for other studies.

\subsection{The threshold-free $\lambda_{DS}$ function}

As discussed before, the common issue in all previous works of classifying cluster dynamical states with either single or multiple dynamical parameters is that the thresholds for these parameters are chosen arbitrarily. 
In order to overcome such issue, we assume that a mass-complete cluster sample at $z=0$ can be roughly separated into dynamically relaxed and unrelaxed from the DMO simulations. The DMO instead of hydro simulations are chosen because the DMO simulations are very robust, different codes give very small difference \citep[e.g.][]{Sembolini2016,Sembolini2016b} -- unlike hydro-simulations of which the internal structures can be significantly altered due to different baryon models \citep[e.g.][]{Cui2016,Elahi2016}. Here, we introduce a new relaxation parameter, $\lambda_{DS}$ here, which is a modification version of \autoref{E7}:

\begin{equation}
    \lambda_{DS}= \sqrt{\frac{3}{(a\times{\Delta_{r}})^{2}+(b\times{f_{s}})^{2}+|1-\eta|^{2}}}.
	\label{E8}
\end{equation}

Instead of using prefactors of $\Delta_{r}$, $f_{s}$ and $|1-\eta|$ terms representing their thresholds of cluster dynamical states, our new $\lambda_{DS}$ completely remove these threshold with its prefactors $a$ and $b$ fitting-determined from our assumption -- $\lambda_{DS}$ has a Double-Gaussian distribution over the DMO clusters. Since we only care about the distribution pattern of $\lambda_{DS}$, instead of their absolute values, the only important thing in \autoref{E8} is the relative contributions from $\Delta_{r}$, $f_{s}$ and $|1-\eta|$. Therefore, we can arbitrarily set one prefactor from one term in the denominator to be 1, i.e here we choose |1-$\eta$| term to have unit prefactor. The prefactors $a$ and $b$ can then be determined by fitting the distributions of $\lambda_{DS}$ with a double-Gaussian function, and finding out the pair of $a$ and $b$ which can give the “best” Double-Gaussian distribution. After the two families of clusters are fitted, it is naturally to use the crossing point of the two Gaussian curves as the threshold for separating the cluster into relaxed and unrelaxed.

\subsection{Determine coefficients $a$ and $b$}
To determine the $a$ and $b$ parameters, a two dimensional array is made. Each element in this array is a pair of trial $a$ and $b$, with the separation of 0.01 between neighbouring elements. For each pair of $a$ and $b$, the distribution of $\lambda_{DS}$, estimated by \autoref{E8}, is used to fit a double-Gaussian function with the free parameters $c_{1}$, $c_{2}$, $\mu_{1}$, $\mu_{2}$ and $\sigma_{1}$, $\sigma_{2}$ to be determined by

\begin{equation}
    f(x) = c_{1}e^{-\frac{(x-\mu_1)^{2}}{2\sigma_1^2}}+c_{2}e^{-\frac{(x-\mu_2)^{2}}{2\sigma_2^2}}.
    \label{E9}
\end{equation}

Several constraining criteria are made to select $a$ and $b$ values. 

Firstly, the list of $\lambda_{DS}$ must {\it not} pass the Shapiro-Wilk test\citep{shapiro}, which aims at testing normal distribution with a single peak. $shapiro()$ function can be directly imported from $scipy.stats$, and it will return an indicator called $P_{value}$ when acting on a list-like object. The distribution is normal if its $P_{value}$ is greater than 0.05. Because we are looking for a double-Gaussian distribution, which shouldn't be normal. Therefore, we require $P_{value} < 0.05$ for the list of $\lambda_{DS}$.

Secondly, 6 parameters ($c_1$, $c_2$, $\mu_1$, $\mu_2$, $\sigma_1$, $\sigma_2$) are obtained from fitting the double-Gaussian distribution with each pair of trial $a$ and $b$, which describe the shapes of two Single-Gaussian functions, see \autoref{E9}. In a well-behaved Double-Gaussian function, the two peaks are expected to have similar heights, similar widths, and be relatively well separated. Therefore, additional constraining criteria are set to be $|c_1 - c_2|< 10$ (similar heights), $|\mu_1 - \mu_2| > 0.2$ (large separation, in log10 scale), and $|\sigma_1-\sigma_2| < 0.05$ (similar widths). 

These four constraints are summarised below, as:

\begin{equation}
    P_{value} < 0.05
    \label{E10}
\end{equation}

\begin{equation}
    |c_1 - c_2| < 10
    \label{E11}
\end{equation}

\begin{equation}
    |\mu_1 - \mu_2| > 0.2
    \label{E12}
\end{equation}

\begin{equation}
    |\sigma_1 - \sigma_2| < 0.05
    \label{E13}
\end{equation}

Lastly, with all qualified pairs of $a$ and $b$ selected, the best candidate is the one which has the least square fitting error $E$. The square fitting error can be calculated from the real distribution of $\lambda_{DS}$ with the predicted values from the fitting result, which is defined as

\begin{equation}
    E = \sum\limits_{i=n}^{N}(y_n-f(x_n))^2.
    \label{E14}
\end{equation}

In this case, $y_n$ is the height of the $n^{th}$ bin, which represents the number of clusters with $\lambda_{DS}$ values within the range of the bin centring at $x_n$. $N$ is the total number of bins and $f(x_n)$ is the fitting value of the Double-Gaussian function at $x_n$.

\begin{figure}
     \centering
     \includegraphics[width=\columnwidth]{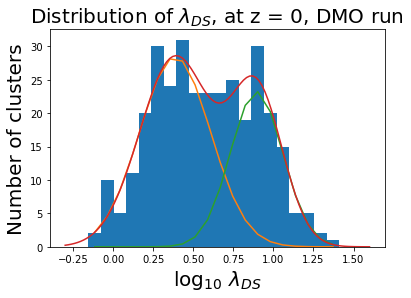}
     \caption{Distributions of the relaxation parameter, $\log_{10} \lambda_{DS}$, for the mass-complete cluster sample from the DMO run, at redshift z = 0. The best fitting parameters are $a = 7.30$ and $b = 0.30$. Red line represent the fitted Double-Gaussian distribution. The two single-Gaussian functions are represented by orange and green line.}
     \label{F1}
\end{figure}

The best fitting parameters for the DMO run are $a = 7.30$ and $b = 0.30$. The $\lambda_{DS}$ distribution with the fitting results is plotted in \autoref{F1}. The best fit parameters for the Double-Gaussian function are listed in \autoref{T1}. The value of $a$ is much larger than $b$, which indicates that the $\lambda_{DS}$ function place far more emphasis on centre-of-mass offset than both the sub-halo mass fraction and the virial ratio. This implies that centre-of-mass offset $\Delta_{r}$ plays an dominated role in $\lambda_{DS}$. We also notice that there is a linear correlation between $\Delta_{r}$ and $f_{s}$, so the small contribution from sub-halo mass fraction provides additional information going beyond the linear correlation.

We specially note here that although the method is reliable, the fitting parameters $a$ and $b$, thus the threshold for separating relaxed and unrelaxed clusters, can be sample dependent, i.e. reducing or increase the minimum cluster mass in the mass-complete sample,  $a$ and $b$ can be changed slightly. However, we are limited to our sample in this study, and as long as we are consistent, our results won't quantitatively change. This is because these key quantities: $f_s$, $\Delta$ and $\eta$ are all physical, which should not change along mass and redshift. Therefore, the same $\lambda_{DS}$ classified as relaxed at $z=0$ or for cluster with higher mass, should be equally relaxed at high $z$ or a lower mass. Further investigation regarding the changes of $a$ and $b$ at with different samples requires much large simulation, we leave it for a later study.

\begin{table}
\centering
\begin{tabular}{| l | l | l | l | l | l | l | l |}
\hline
 $P_{value}$ & $c_1$ & $c_2$ & $\mu_1$ & $\mu_2$ & $\sigma_1$ & $\sigma_2$ & $E$ \\
\hline
0.01 & 35.35 & 30.38 & 0.34 & 0.80 & 0.18 & 0.13 & 115.20 \\
\hline
\end{tabular}
\caption{$P_{value}$, coefficients c, $\mu$, $\sigma$, square error $E$ for the fitting result with the best parameters: $a = 7.30$ and $b = 0.30$.}
\label{T1}
\end{table}

\subsection{The threshold for separating relaxed and unrelaxed clusters}

The Double-Gaussian distribution of $\lambda_{DS}$ naturally avoid the arbitrary choice of the threshold for separating dynamically relaxed and unrelaxed clusters. The threshold is defined as the x coordinate of the crossing point of two Single-Gaussian functions, see \autoref{F1}. The threshold value is $\lambda_{DS} = 3.424$ in normal scale.

\begin{figure}
     \centering
     \begin{subfigure}{0.4\textwidth}
         \centering
         \includegraphics[width=\columnwidth]{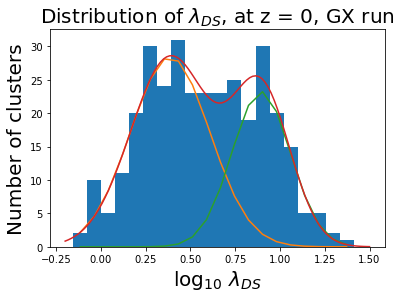}
         \caption{}
         \label{F2.1}
     \end{subfigure}
     \newline
     \centering
     \begin{subfigure}{0.4\textwidth}
         \centering
         \includegraphics[width=\columnwidth]{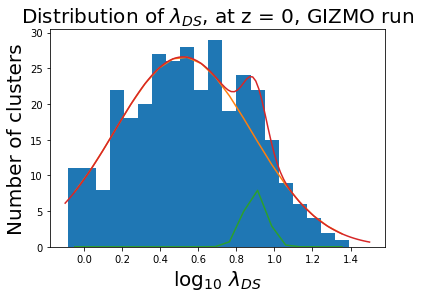}
         \caption{}
         \label{F2.2}
     \end{subfigure}
        \caption{Distributions of relaxation parameter, $\lambda_{DS}$, in log10 scale (the same as \autoref{F1}) for 324 clusters at redshift $z = 0$, in GX run (a) and GIZMO run (b). Red lines represent Double-Gaussian fit, single-Gaussian fits are represented by orange and green lines in each plot.}
        \label{F2}
\end{figure}

In different hydro-dynamical simulations, different baryonic models are used, which can result in different best-fitted Double-Gaussian functions. For the simplicity in this investigation, and in order to highlight the changes due to different baryon models, we apply the same fitting results from DMO fitting as the baseline, i.e. with $a$ = 7.3, $b$ = 0.3, to calculate $\lambda_{DS}$ in GX and GIZMO runs. The distributions of $\lambda_{DS}$ in GX and GIZMO runs are plotted in \autoref{F2}, along with the fitting results. Note that the distribution of $\lambda_{DS}$ from GIZMO can not be fitted to double-Gaussian and we never expect that as the baryon models will change cluster dynamical state. Although the distribution of $\lambda_{DS}$ from GX can be fitted to double-Gaussian, there is a shift of the threshold value compared to the DMO result. Nevertheless, applying the same parameters with threshold, we can examine the effects of baryons. For example, with the same threshold, $\lambda_{DS} = 3.424$, applied to GX and GIZMO run, we find that $151/171/170$ clusters are classified as relaxed clusters in DMO/GX/GIZMO run. It looks that hydro-simulations with baryon model tend to increase the number of relaxed clusters. More details will be presented in \S\autoref{subs:dsb}.

\subsection{The relationship between $\lambda_{DS}$ and $\chi_{DS}$} 

\begin{figure}
     \centering
     \includegraphics[width=\columnwidth]{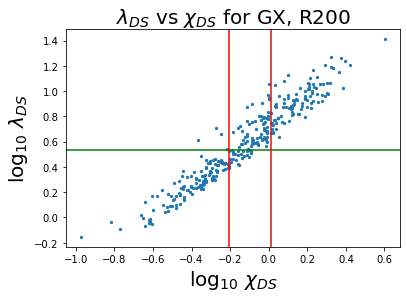}
     \caption{$\lambda_{DS}$ vs $\chi_{DS}$ in logarithm scale for 324 clusters, at z = 0, in GX run, which \citet{300Haggar2020} used. Two red vertical lines represent the threshold on $\chi_{DS}$ for relaxed clusters and unrelaxed clusters, which are $\chi_{DS} = 1.030$ and $\chi_{DS} = 0.619$ respectively. The green vertical line represents the threshold on $\lambda_{DS}$, which is $\lambda_{DS} = 3.424$.}
     \label{F11}
\end{figure}

This new relaxation parameter, $\lambda_{DS}$, is compared with the corresponding old relaxation parameter, $\chi_{DS}$, in \autoref{E7}. Based on the value of $\chi_{DS}$, \cite{300Haggar2020} split the sample as relaxed clusters ($\chi_{DS}$ > 1.030), unrelaxed clusters ($\chi_{DS}$ < 0.619) and intermediate with 0.619 < $\chi_{DS}$ < 1.030. The two thresholds from \cite{300Haggar2020} are represented by two red vertical lines in \autoref{F11}. In our work, a single threshold, $\lambda_{DS}$ = 3.424 is determined from a systematical way, which is represented by the green horizontal line. The correlation between $\lambda_{DS}$ and $\chi_{DS}$ is almost linear, and the most clusters classified as relaxed or unrelaxed by \cite{300Haggar2020} have a similar classification with our parameter. This means although our new relaxation parameter $\lambda_{DS}$ adjusts the relative contributions between dynamical parameters $\eta$, $f_{s}$ and $\Delta_{r}$ to re-scale and re-distribute the old relaxation parameter $\chi_{DS}$, it is still monotonic correlated with old one and almost doesn't qualitatively change the results from previous work on classifying cluster dynamical states. Whilst, $\lambda_{DS}$ can provide a single and non-arbitrary threshold from its special Double-Gaussian distribution, which makes the classification of cluster dynamical states straightforward and clear.

\section{Results} \label{s:results}

\subsection{The baryon effect on the cluster dynamical state}
\label{subs:dsb}

\begin{figure}
     \centering
    
    \begin{subfigure}{0.4\textwidth}
        \centering
        \includegraphics[width=\columnwidth]{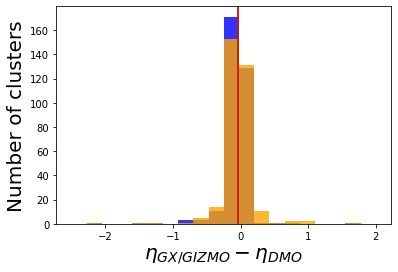}
        \caption{}
        \label{F3.1}
    \end{subfigure}
    \newline
    \centering
    \begin{subfigure}{0.4\textwidth}
        \centering
        \includegraphics[width=\columnwidth]{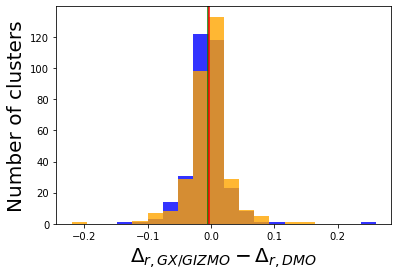}
        \caption{}
        \label{F3.2}
    \end{subfigure}
    \newline
    \centering
    \begin{subfigure}{0.4\textwidth}
        \centering
        \includegraphics[width=\columnwidth]{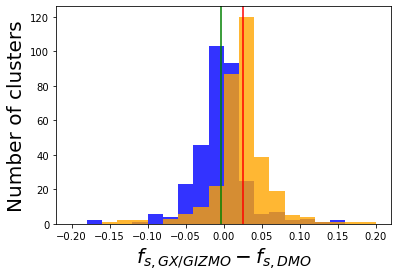}
        \caption{}
        \label{F3.3}
    \end{subfigure}
    \newline
    \centering
    \begin{subfigure}{0.4\textwidth}
        \centering
        \includegraphics[width=\columnwidth]{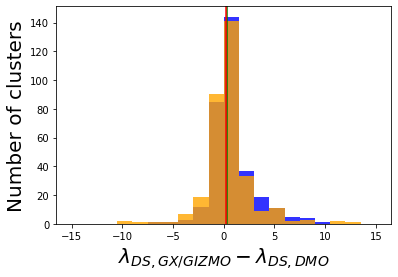}
        \caption{}
        \label{F3.4}
    \end{subfigure}
    \caption{Distributions of the relative differences for (from top to bottom) $\eta$, $\Delta_{r}$, $f_s$, $\lambda_{DS}$ between GX (blue histograms) or GIZMO (orange histograms) run and DMO run. Green (red) vertical lines represent the median number of differences for GX (GIZMO) run.}
    \label{F3}
\end{figure}

\begin{table}
\centering
\begin{tabular}{| l | l | l | l | l |}
\hline
  & $\eta$ & $\Delta_{r}$ & $f_s$ & $\lambda_{DS}$ \\
\hline\hline
GX - DMO & -0.039(0.232) & -0.005(0.030) & -0.003(0.036)& 0.289(2.471)  \\
\hline
GIZMO - DMO & -0.034(0.256) & -0.003(0.036) & 0.025(0.036)& 0.180(2.554)  \\
\hline
DMO & 1.155 & 0.066 & 0.143& 3.017 \\
\hline
\end{tabular}
\caption{The median numbers of the differences of (from left to right) $\eta$, $\Delta_{r}$, $f_{s}$ and $\lambda_{DS}$ between GX run (the first row)/GZIMO run (the second row) and DMO run (numbers in brackets are standard errors), The third row displays the median numbers of each parameter in DMO run.}
\label{T2}
\end{table}

We further investigate the baryon effect on dynamical parameters $\eta$, $f_s$, $\Delta_{r}$ and $\lambda_{DS}$. We matched the corresponding clusters from different runs at $z$ = 0. For each cluster, the differences of these parameters between hydro-dynamical simulation (GX or GIZMO) and DMO simulation are shown in \autoref{F3}. Distributions of these differences are plotted in histograms. For each parameter, the median numbers of these differences are used to quantify the baryon effect. Those median numbers and standard errors are marked as vertical lines in \autoref{F3} and listed in \autoref{T2}. Main results of baryon effects are discussed below: 

\begin{itemize}
\item $\eta$ from the GX run and the GIZMO run are reduced by about 2\% compared to the median value from DMO run. The differences distribution between the two hydro-dynamical simulations is very small, which gives the insight that the effect on $\eta$ depends weakly on baryon models. \cite{cui_dynamical_2017} showed a similar result on the weakly-model-dependent effect of the decrease in $\eta$, but with a larger difference, about 10\%, for CSF run and AGN run. Here CSF run referred to a hydro-dynamical simulation ignoring the AGN feedback, and AGN run includes AGN feedback. They also concluded that the ratio between $\eta$ from hydro-dynamical run and $\eta$ from DMO run shows no dependence on cluster mass.

\item Standard deviations of the differences between $\Delta_r$ from GX/GIZMO run and from DMO run are comparable to the scale of the median number of $\Delta_r$ from DMO run, which shows the scattering distribution of $\Delta_{r}$ in hydro-dynamical simulation, in agreement with the result in \cite{cui_dynamical_2017}. This is the consequence of that the position of substructure can be largely affected by baryons. However, the average amount of change for all clusters is small, which behaves as a small decrease about 5\% compared to DMO run. This could be mainly caused by the central galaxy formation which deeps the potential and increases the halo concentration. Thus, more weights are contributed from the central region. 

\item Comparing to the DMO value, $f_s$ increases by 17\% in GIZMO run. This is in agreement with the result on \cite{cui_dynamical_2017}: their CSF run increases the $f_s$ by 40\% for higher cluster mass, and by 20\% for clusters with lower masses.

However, the median change of $f_s$ in GX run is negligible, smaller than 5\%. The difference between $f_{s}$ from GX run and GIZMO run should come from the feedback models that control the galaxy formation in these less massive substructures. Through the comparison of satellite stellar mass function in Cui et al. 2021 (in prep.), it is clear that the satellite stellar mass function from the GIZMO run agrees better with the observation results at lower galaxy mass than GX which is about 5 times lower.

\item $\lambda_{DS}$ in GX (GIZMO) run is 9 (6) per cent higher than in DMO run, which indicates a weak the baryon-model dependence of $\lambda_{DS}$. This is not surprising as the baryon models have a weak influence on the individual parameters.
\end{itemize}

\subsection{Dynamical state and cluster mass accretion history} \label{subs:mah}
\begin{figure}
     \centering
    
    \begin{subfigure}{0.4\textwidth}
        \centering
        \includegraphics[width=\columnwidth]{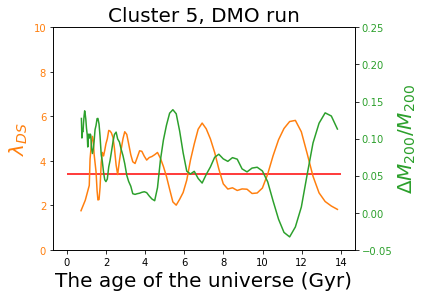}
        \caption{}
        \label{F4.1}
    \end{subfigure}
    \newline
    \centering
    \begin{subfigure}{0.4\textwidth}
        \centering
        \includegraphics[width=\columnwidth]{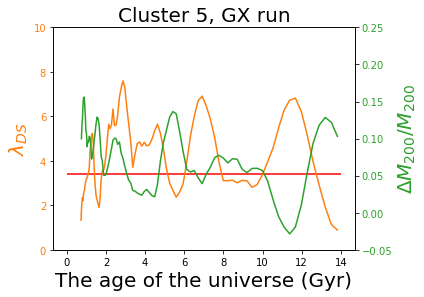}
        \caption{}
        \label{F4.2}
    \end{subfigure}
    \newline
    \centering
    \begin{subfigure}{0.4\textwidth}
        \centering
        \includegraphics[width=\columnwidth]{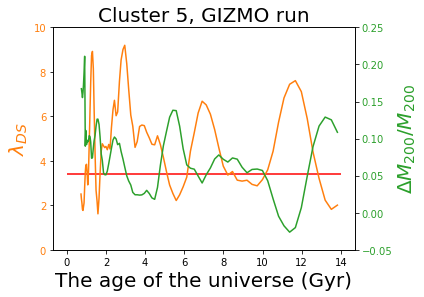}
        \caption{}
        \label{F4.3}
    \end{subfigure}
    \caption{Evolution tracks of $\lambda_{DS}$ (orange) and $\frac{\Delta M_{200}}{M_{200}}$ (green) over time for the 5th cluster, in (a) DMO run; (b) GX run; (c) GIZMO run. The red horizontal line represents the threshold $\lambda_{DS} = 3.424$. The region above this line represents the cluster in a dynamically relaxed state.}
    \label{F4}
\end{figure}

It is clear that the cluster dynamical state changes are caused by the accretion of mass, especially in the case of major merger events. This is also revealed from the three key items in \autoref{E8}.
However, it is unclear how significant the cluster dynamical state can be altered and how long the cluster will return to relaxed state after a merger event. In this section, we will try to quantify the relationship between cluster dynamical states and the mass changes, and investigate the relaxation time scale -- from the beginning of a disturbance to the final relaxed state (see more details in the following section). In \autoref{F4}, we illustrate the evolution tracks of $\lambda_{DS}$ and $\frac{\Delta M_{200}}{M_{200}}$ over time for one arbitrary example cluster, where the variation $\Delta M_{200}$is estimated by $M_{200}$ in the snapshot i minus $M_{200}$ in snapshot i-1. The original behavior of the evolution track of $\lambda_{DS}$ is highly jagged because of the frequent mergers along with difficulties in correctly tracking the progenitors in the simulation, so the function $savagol\_filter$ from $SciPy.signal$ is applied to smooth the evolution curve (see scipy.org). In this function, the length of the filter window is set to be 11 data points, and the order of the polynomial to fit the sample is set to be 3.

\subsection{The cluster relaxation time scale} \label{subs:rts}
\begin{figure}
	\includegraphics[width=\columnwidth]{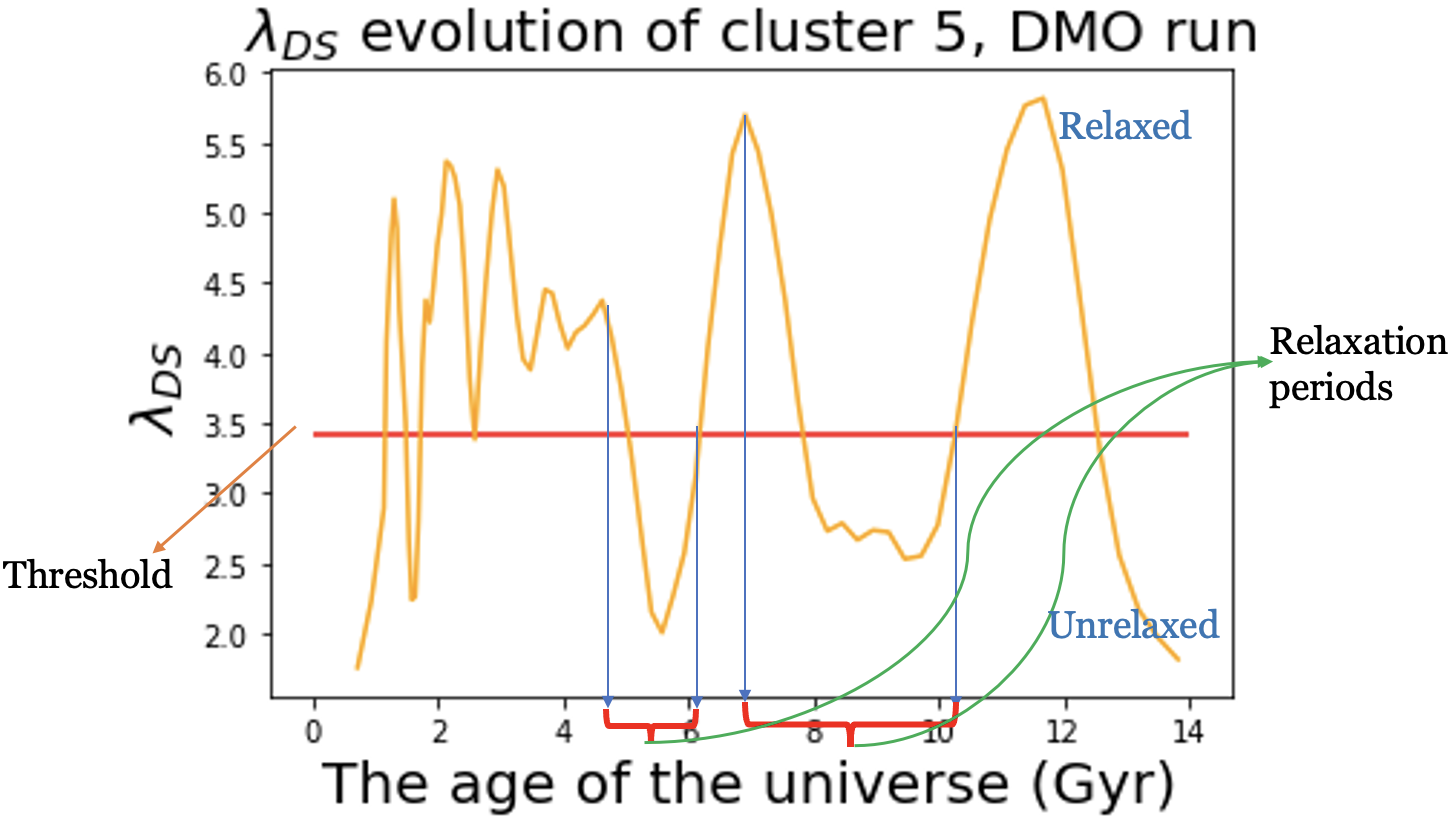}
    \caption{$\lambda_{DS}$ evolutionary track for the 5th cluster in DMO run. The red brackets labels two relaxation periods identified in the evolutionary process of the cluster. The red horizontal line represent the threshold of cluster dynamical states, above and below which are relaxed and unrelaxed states respectively.}
    \label{F5}
\end{figure}

In order to investigate the evolution of cluster dynamical states, we define the "relaxation period" to describe the time taken by a cluster to evolve from relaxed state to unrelaxed state and then return to relaxed states. As shown in \autoref{F5}, one relaxation period starts with the local maximum of $\lambda_{DS}$ above the threshold before decreasing, after which the $\lambda_{DS}$ of cluster continues decreasing until reach some local minimum below the threshold. Then, the relaxation period ends with the first crossing point between $\lambda_{DS}$ evolutionary track and the threshold, through which the cluster return to a relaxed state again. Note that we exclude the evolution track in the very beginning 4 Gyrs. This is because the halo is still very small and its dynamical state can be dramatically changed due to very frequent merging events. Our definition of this relaxation time scale is very similar to the merger time which is defined in Contreras-Santos et al. 2021. We share the same initial point to mark the start of relaxation time scale. However, Contreras-Santos et al. 2021 require the cluster returns to a following peak of the dynamical relaxation parameter for the end instead of the crossing of the threshold (our case). Besides that, they used $\chi_{DS}$ parameter to quantify the cluster dynamical state which is very similar to our $\lambda_{DS}$ as shown in \autoref{F11}. Therefore, we expect a similar scale between their merger time and our relaxation time. It worth noting that their studies focus on major merger events ($\Delta_M/M \geq 0.5$), while we will provide a more statistical view of the relaxation time scale.

\begin{figure}
    \centering
    \begin{subfigure}{0.4\textwidth}
        \centering
        \includegraphics[width=\columnwidth]{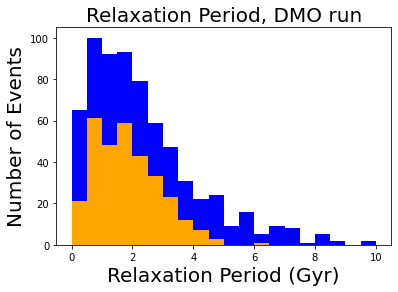}
        \caption{}
        \label{F6.1}
    \end{subfigure}
    \newline
    \centering
    \begin{subfigure}{0.4\textwidth}
        \centering
        \includegraphics[width=\columnwidth]{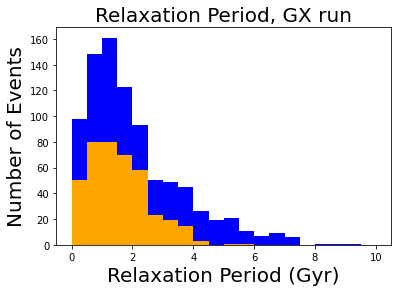}
        \caption{}
        \label{F6.2}
    \end{subfigure}
    \newline
    \centering
    \begin{subfigure}{0.4\textwidth}
        \centering
        \includegraphics[width=\columnwidth]{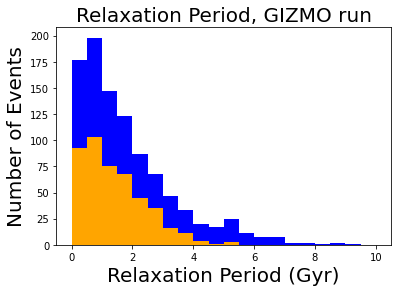}
        \caption{}
        \label{F6.3}
    \end{subfigure}
    \caption{Relaxation period distribution for all clusters, in (a) DMO run; (b) GX run; (c) GIZMO run. Blue bins represent all relaxation periods, orange bins represent relaxation periods with only one $\frac{\Delta M_{200}}{M_{200}}$ peak inside.}
    \label{F6}
\end{figure}

As shown in \autoref{F5}, one cluster can have more than one relaxation periods during its evolution process. The distributions for relaxation periods for clusters in samples are shown in \autoref{F6}. The relaxation time scale is quantified as the median number of relaxation periods, which are 1.9 (1.8) Gyr, 1.6 (1.6) Gyr and 1.4 (1.6) Gyr for DMO run, GX run and GIZMO run respectively, the numbers inside brackets are standard deviations.

\subsubsection{Connection to the halo mass changes}
Relaxation time scale provides useful information about the evolution of cluster dynamical states, especially when connecting with the merger events. As mentioned above, it is intuitively correlated to the mass accretion history of cluster. 

In order to investigate such correlation, we first select out those relaxation periods each with only one $\frac{\Delta M_{200}}{M_{200}}$ peak inside, and make the scatter plots of relaxation time period vs the maxima of fractional halo mass change, $\frac{\Delta M_{200}}{M_{200}}$. However, no direct correlation can be found in this scattering plot. Then, we find that the relaxation period, after normalized with a dynamical time scale,$t_{dyn}$, $t_{relax}/t_{dyn}$ then shows a moderate linear correlation with $\frac{\Delta M_{200}}{M_{200}}$ peak. The dynamical time scale $t_{dyn}$ is define as  

\begin{equation}
    t_{dyn} = (\frac{R_{vir}^{3}}{GM_{vir}})^{1/2},
    \label{E15}
\end{equation}

where $R_{vir}$ and $M_{vir}$ are virial radius and virial mass respectively. Here, we simply adopt $R_{200}$ as the virial radius and $M_{200}$ as the virial mass.

From \autoref{E15}, it is easy to show that the dynamical time scale, $t_{dyn}$, is only a function of critical density, $\rho_{c}$, which solely depends on redshift $z$. Hence, the dynamical time scale can be determined only with a given redshift. In this study, $t_{dyn}$ is determined from the redshifts at which the relaxation periods start.

\begin{figure}
    \centering
    \begin{subfigure}{0.4\textwidth}
        \centering
        \includegraphics[width=\columnwidth]{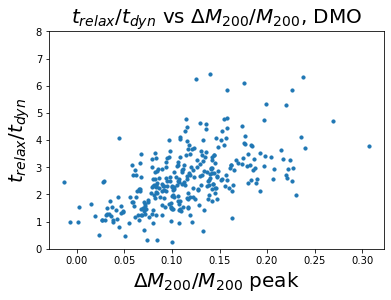}
        \caption{}
        \label{F7.1}
    \end{subfigure}
    \newline
    \begin{subfigure}{0.4\textwidth}
        \centering
        \includegraphics[width=\columnwidth]{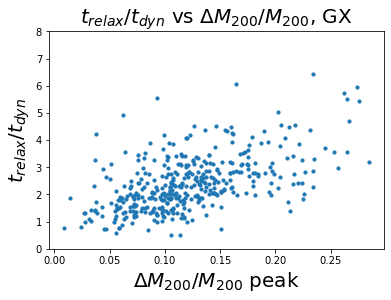}
        \caption{}
        \label{F7.2}
    \end{subfigure}
    \newline
    \begin{subfigure}{0.4\textwidth}
        \centering
        \includegraphics[width=\columnwidth]{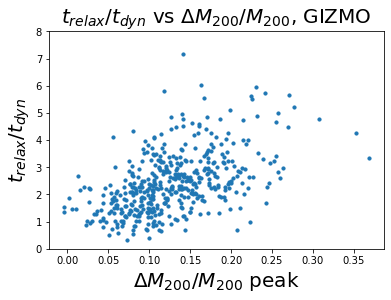}
        \caption{}
        \label{F7.3}
    \end{subfigure}
    \caption{$\frac{t_{relax}}{t_{dyn}}$ vs $\frac{\Delta M_{200}}{M_{200}}$, for relaxation periods with single $\frac{\Delta M_{200}}{M_{200}}$ peak, for (a) DMO run; (b) GX run; (c) GIZMO run.}
    \label{F7}
\end{figure}

The scatter plots of relaxation period/dynamical time scale vs $\frac{\Delta M_{200}}{M_{200}}$ are shown in \autoref{F7}. The Pearson correlation coefficient is 0.57/0.55/0.51 for DMO/GX/GIZMO run, which indicates a moderate correlation. Including more halo properties may give a better correlation, we retain that for a future study. Note that, as have discussed in \autoref{subs:rts}, it is not surprise to see a similar distribution of $t_{relax/t_{dyn}}$ in \autoref{F7} compared to the Fig. 3 in Contreras-Santos et al. 2021. 

Then, we fitted these scatter plots with a linear function:

\begin{equation}
    \frac{t_{relax}}{t_{dyn}} = k\times\frac{\Delta M_{200}}{M_{200}}+h
    \label{E16}
\end{equation}

and we obtain $k$ = 12.047/10.706/9.794 and $h$ = 1.169/1.111/1.183 from DMO/GX/GIZMO run, respectively. For the two hydro-dynamical simulations, we calculate their mean square fitting errors by \autoref{E14} and compare them with the mean square errors from DMO fitting function, i.e. data are from hydro-dynamical run, but predictions are made with \autoref{E16} with parameters k and h yielded from DMO fitting. For GX run, the mean square error from DMO fitting ($k$ = 12.047, $h$ = 1.169) is 0.36, and that from GX fitting ($k$ = 10.706, $h$ = 1.111) is 0.34. For GIZMO run, the mean square error from DMO fitting is 0.34, and that from GIZMO fitting ($k$ = 9.794, $h$ = 1.183) is 0.33. The differences in mean square fitting errors from DMO fitting and hydro-dynamical fitting are small in both cases. Therefore, we simply use the values of k and h from the DMO fitting for all three simulations.

\begin{figure}
    \centering
    \begin{subfigure}{0.4\textwidth}
        \centering
        \includegraphics[width=\columnwidth]{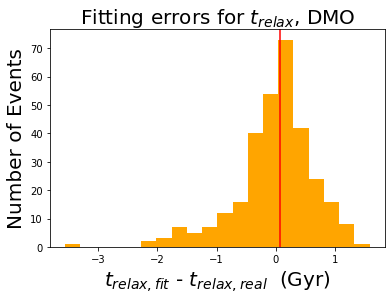}
        \caption{}
        \label{F8.1}
    \end{subfigure}
    \newline
    \begin{subfigure}{0.4\textwidth}
        \centering
        \includegraphics[width=\columnwidth]{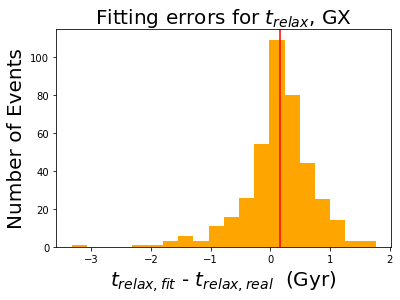}
        \caption{}
        \label{F8.2}
    \end{subfigure}
    \newline
    \begin{subfigure}{0.4\textwidth}
        \centering
        \includegraphics[width=\columnwidth]{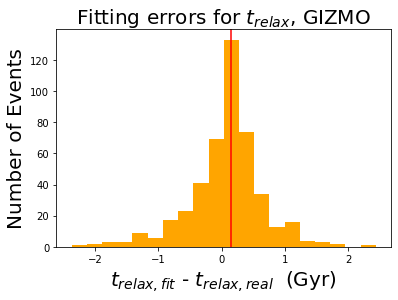}
        \caption{}
        \label{F8.3}
    \end{subfigure}
    \caption{Fitting errors of relaxation periods with single $\frac{\Delta M_{200}}{M_{200}}$ peak inside the duration, for (a) DMO run; (b) GX run; (c) GIZMO run. Red vertical line represents the median number of fitting errors.}
    \label{F8}
\end{figure}

The distributions of fitting errors for relaxation periods with single $\frac{\Delta M_{200}}{M_{200}}$ peak inside are showed in \autoref{F8}. Most errors between predicted and real relaxation periods (89\%/91\%/89\% for DMO/GX/GIZMO run) are less than $\sim$0.5 Gyr, which are considerably less than the median length of those relaxation periods with single $\frac{\Delta M_{200}}{M_{200}}$ peak, which is 1.847/1.577/1.413 Giga years for DMO/GX/GIZMO run. The median fitting errors of three distributions are close to 0, slightly deviate towards a positive direction.

\begin{figure}
    \centering
    \begin{subfigure}{0.4\textwidth}
        \centering
        \includegraphics[width=\columnwidth]{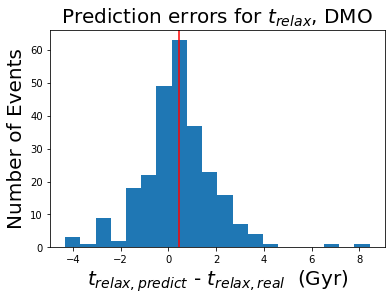}
        \caption{}
        \label{F9.1}
    \end{subfigure}
    \newline
    \begin{subfigure}{0.4\textwidth}
        \centering
        \includegraphics[width=\columnwidth]{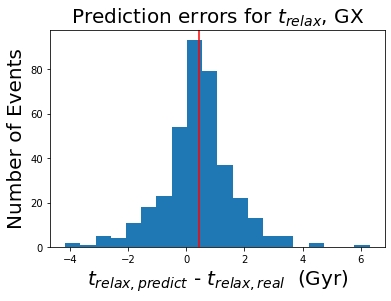}
        \caption{}
        \label{F9.2}
    \end{subfigure}
    \newline
    \begin{subfigure}{0.4\textwidth}
        \centering
        \includegraphics[width=\columnwidth]{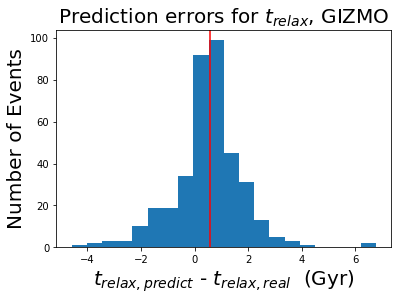}
        \caption{}
        \label{F9.3}
    \end{subfigure}
    \caption{Prediction errors of relaxation periods with multiple $\frac{\Delta M_{200}}{M_{200}}$ peaks inside the duration, for (a) DMO run; (b) GX run; (c) GIZMO run. Red vertical lines represent the median prediction errors.}
    \label{F9}
\end{figure}

Given the linear correlation shown in \autoref{F7}, it is not surprising to see such a relatively small fitting error. To verify this fitting function, we adopt it for making predictions of the relaxation periods with more than one $\frac{\Delta M_{200}}{M_{200}}$ peaks by simply linear summation of contributions from all $\frac{\Delta M_{200}}{M_{200}}$ peaks:

\begin{equation}
    t_{relax} = t_{dyn}\times\sum\limits_{i=1}^{n}(12.047\times\frac{\Delta M_{200,i}}{M_{200,i}}+1.169),
    \label{E17}
\end{equation}

where $t_{dyn}$ is calculated by the redshift at which the relaxation period starts, and n represents the total number of $\frac{\Delta M_{200}}{M_{200}}$ peaks that happen within the relaxation period. Note that a different $t_{dyn}$ for each peak may give a better prediction. The distributions of fitting errors for those relaxation periods with multiple peaks inside are showed in \autoref{F9}. Most errors (82\%/88\%/88\% for DMO/GX/GIZMO run) are less than $\sim$2 Gyrs. However, the median numbers of these distributions deviate towards the positive direction ($\la 0.5$ Gyr), which means that \autoref{E17} slightly overestimate the length of relaxation period.

\begin{figure}
     \centering
     \includegraphics[width=\columnwidth]{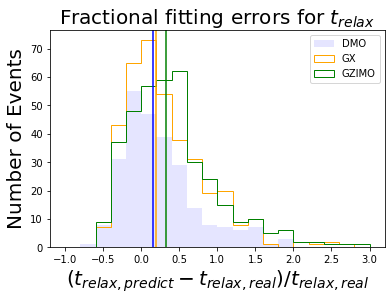}
     \caption{Distributions of the relative prediction errors, ($t_{relax, predict}$-$t_{relax, real}$)/$t_{relax,real}$, for clusters in DMO(filled blue bins), GX(unfilled orange line) and GIZMO(unfilled green line) run. Blue/Orange/Green vertical line represents the median number of prediction error from DMO/GX/GIZMO run.}
     \label{F10}
\end{figure}

The fractional fitting error distributions for DMO, GX and GIZMO run are plotted altogether in \autoref{F10}. The histograms are not normalized, the total number of relaxation periods in two hydro-dynamical runs are significantly larger than that in DMO run, which implies an increased merger events by the baryon effect. 81.3\%/74.1\%/71.2\% of fractional errors in DMO/GX/GIZMO run are less than 0.6. In agreement with the behaviors in absolute error distributions, all fractional error distributions deviate towards positive direction. The deviation of the median number of fractional fitting error is strongest in GIZMO run, and the median number in GX run also has a larger deviation than that in DMO run.

\section{Conclusions and discussions} \label{s:conc}
In this work, we use the mass-complete cluster sample from \theth\ to study the cluster dynamical states and proposed a new parameter $\lambda_{DS}$ to classify the clusters into dynamical relaxed and unrelaxed without a manually set threshold. Benefited from the different runs (DMO, GX and GIZMO) within this project, we are also able to investigate the baryon effect on the cluster dynamical state. Furthermore, we define a relaxation time scale and connect it to the halo mass changes. Main findings are summarised below:

\begin{itemize}

\item Based on the relaxation parameter $\chi_{DS}$ in \cite{300Haggar2020}, a new threshold-free function of $\lambda_{DS}$ is proposed to classify cluster dynamical states, which is 

\begin{equation}
    \lambda_{DS}= \sqrt{\frac{3}{(7.30\times{\Delta_{r}})^{2}+(0.30\times{f_{s}})^{2}+|1-\eta|^{2}}}
	\label{E18}
\end{equation}

The threshold distinguishing relaxed and unrelaxed state is naturally set by the double-Gaussian fitting of the $\lambda_{DS}$ distribution. At redshift z = 0, 151/171/170 clusters of all 324 clusters are classified to be dynamically relaxed in DMO/GX/GIZMO run. The $\lambda_{DS}$ parameter is linearly correlated to $\chi_{DS}$ parameter, and it preserves the classification results based on $\chi_{DS}$.

\item Including baryons to simulations can slightly reduce the virial ratio $\eta$, which is 2\% lower in GX and GIZMO run compared to DMO run. 

Baryonic effect results in the scattering distribution of the center of mass offset, $\Delta_{r}$, the standard deviation of the difference between $\Delta_{r}$ from GX/GIZMO run and DMO run is large (more than 50\%) compared to the scale of $\Delta_{r}$ from DMO run.

Sub-halo mass fraction $f_s$ is 17\% higher in the GIZMO run than in the DMO run, while the GX run is about 2 per cent lower.

In combination, the $\lambda_{DS}$ in GIZMO run is 3\% lower than in GX run which has about 10 per cent higher value than the DMO run. Therefore, more relaxed clusters are presented in the hydrodynamic simulations. Nevertheless, the baryons play a weak role in altering the cluster dynamical state.

\item The median number of relaxation periods (the time taken by a cluster to evolve from most relaxed state to unrelaxed state and then return to relaxed state) also regarded as relaxation time scale, is 1.913/1.610/1.419 for DMO/GX/GIZMO run, respectively. 


\item The relaxation period is correlated to cluster mass accretion history. For relaxation periods with single $\frac{\Delta M_{200}}{M_{200}}$ peak inside, a moderate linear correlation is observed, which is described as 
\begin{equation}
    \frac{t_{relax}}{t_{dyn}} = 12.047\times\frac{\Delta M_{200}}{M_{200}}+1.169.
    \label{E19}
\end{equation}

In general case, the length of relaxation periods can be predicted from the heights of $\frac{\Delta M_{200}}{M_{200}}$ peaks with
\begin{equation}
    t_{relax} = t_{dyn}\times\sum\limits_{i=1}^{n}(12.047\times\frac{\Delta M_{200,i}}{M_{200,i}}+1.169)
    \label{E20}
\end{equation}

with a considerable small error, basically less than 2 Gyrs.
\end{itemize}

As shown in \autoref{F11}, the new proposed $\lambda_{DS}$ is basically linear correlated with the $\chi_{DS}$. So it can be correlated with these observational measured quantities, such as $M$ \citep{cialone_morphological_2018,300DeLuca2021} and $\mathcal{C}$ \citep{300Capalbo2021} parameters.
By applying its threshold from a double-Gaussian fitting, the clusters can be naturally separated into relaxed and unrelaxed. With this single and non-arbitrary classification, it is straightforward to define some time scale to describe the transition rate of dynamical states of a cluster, and such time scale can be determined completely from the features of the evolution track of $\lambda_{DS}$ (see \autoref{F5}), which makes it applicable to be analyzed statistically for large number of clusters, thus evaluate its overall correlation with other observables (e.g fractional mass change of cluster).

In this work, we only impose two constraints on the $\lambda_{DS}$ parameter: having well-behaved  Double-Gaussian distribution over clusters and preserving classification results with $\chi_{DS}$. Meanwhile, the observed linear correlation between sub-halo mass fraction $f_{s}$ and center of mass offset $\Delta_{r}$ is likely to introduce additional degrees of freedom in Double-Gaussian fitting. Therefore, we acknowledge there may be some other values of $a$ and $b$ in \autoref{E8}, or even a different form of function to combine dynamical parameters together which can make $\lambda_{DS}$ satisfy our requirements. In future work, it will be worthy to investigate the potential improvement of the formalism of $\lambda_{DS}$ with some advanced statistical methods.

Although the baryons can affect the cluster properties in different aspects \citep[see][for example]{Cui2016}, the cluster dynamical state seems to be less influenced by the baryons. That is understandable as baryons have the strongest effect at very small scale, while the dynamical state describes the whole dynamical information of the cluster. This is similar to baryon effects on the total cluster mass \citep[see][for example]{Cui2012b,Cui2014}. Agreed with \cite{Zhang2016}, the baryon effect does shirk the cluster's relaxation period, which results in slightly more relaxed clusters in the hydro-dynamical runs. However, different to their ideal case study which doesn't include any baryon processes in two-halos merger event, the hydro-simulated clusters from \theth\ project do not show significantly change in the relaxation time scale. This can be explained as in reality the merger speed and the gas content are relatively low, which is in agreement with their results -- $\sim 70$ per cent reduction in the merger time scale.

Note that our definition of the cluster relaxation time scale is slightly different to the merger time which is widely used in the Semi-Analytical models \citep[for example][]{boylan-kolchin_dynamical_2008,Jiang2008,Jiang2014}. Our definition focus on the overall cluster dynamical state, while the merger time mainly interest in the dynamical friction, for example, a satellite galaxy moving in a dark matter halo. The two time scales are very similar when a major merger happens. Moreover, by using the relation between the cluster dynamical state relaxation time scale with the cluster mass changes in this study, one can roughly predict how long will the cluster will get back to a relaxed state. 

As the merger events can lead to the cluster/galaxy property changes. Contreras-Santos et al. 2021 (in review) using the cluster dynamical changes (similar to our relaxation time scale definition) to define pre- and post-merger phases, found that stellar content of BCGs grows significantly during mergers: the main growth mechanism is the accretion of older stars; there is a burst in star formation induced by the merger. Furthermore, the evolution of the hydrodynamic equillium bias can be also tightly connect to the major mergers (Gianfagna et al. 2021 in prep.). Therefore, though the observed accretion in mass, we can predict the cluster relaxation time scale which can be used to predict the changes of these quantities.

\section*{Acknowledgements}
We kindly thank Alexander Knebe, Ana Contreras, Elena Rasia, Frazer Pearce, Roan Haggar for useful discussions.

This work has received financial support from the European Union's Horizon 2020 Research and Innovation programme under the Marie Sklodowskaw-Curie grant agreement number 734374, i.e. the LACEGAL project.

The simulations used in this paper have been performed in the MareNostrum Supercomputer at the Barcelona Supercomputing Center, thanks to CPU time granted by the Red Espa$\tilde{\rm n}$ola de Supercomputaci$\acute{\rm o}$n. The CosmoSim database used in this paper is a service by the Leibniz-Institute for Astrophysics Potsdam (AIP). The MultiDark database was developed in cooperation with the Spanish MultiDark Consolider Project CSD2009-00064.

WC is supported by the STFC AGP Grant ST/V000594/1. He further acknowledges the science research grants from the China Manned Space Project with NO. CMS-CSST-2021-A01 and CMS-CSST-2021-B01.

\section*{Data Availability}

The data used in this paper is provided by \theth\ project. It is available upon request.



\bibliographystyle{mnras}
\bibliography{reference} 




\appendix

\section{$\lambda_{DS}$ and threshold for R500 data}

\begin{figure}
     \centering
     \includegraphics[width=\columnwidth]{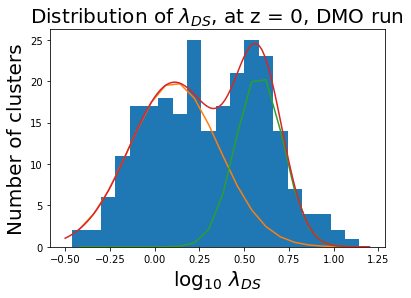}
     \caption{Distributions of the relaxation parameter, $\log_{10} \lambda_{DS}$, for the mass-complete cluster sample from the DMO run, R500, at redshift z = 0. The best fitting parameters are $a = 15.85$ and $b = 1.04$. Red line represent the fitted Double-Gaussian distribution. The two single-Gaussian functions are represented by orange and green line.}
     \label{F12}
\end{figure}

For R500 data, a halo mass cut,$M_{500} = 4.6e14$ is applied to exclude low mass clusters. Then the same method is applied to the left 246 clusters, and the free coefficients for $\lambda$ in eqn \ref{E8} are determined to be a = 15.85 and b = 1.04. The distribution of $\lambda_{DS}$ for R500 is showed in Figure \ref{F12}.

The threshold is $\lambda_{DS} = 2.61$, as the X coordinate of the crossing point of two Single-Gaussian functions. With this threshold, 101 in 246 clusters are classified as relaxed.


\bsp	
\label{lastpage}

\end{document}